\documentstyle[eqsecnum,aps,prd,epsf]{revtex}
\begin{document}
%%%%%%%%%%%%%%%%%%%%%%%%%%%%%%%%%%%%%%%%%%%%%%%%%%%%%%%%%%%%%%%%%%%
\def\pubnum{366/COSMO-83} 
\def\Mark#1{\vspace*{0.5cm} \fbox{{\bf #1}}\\}
\def\NOTE#1{{\bf (*#1*)}}
\def\Isom#1{{\mbox{{\rm Isom}$#1$}}}
\def\UC#1{\widetilde #1}
\def\ME#1{#1'}
%%% Memo %%%
\vspace*{0.0cm}
\def\ma#1{\fbox{ #1}}
\def\memo#1{\footnote{\begin{screen}\ma{Memo}\quad #1\end{screen}}}
%%% footnote %%%
\def\thefootnote{\fnsymbol{footnote}}
%%%%%%%%%%%%%%%%%%%%%%%%%%%%%%%%%%%%%%%%%%%%%%%%%%%%%%%%%%%%%%%%%%%
%\def\Mark{\def\dummy}
%\def\mark{\def\dummy2}
%\def\memo{\def\dummy3}
%%%%%%%%%%%%%%%%%%%%%%%%%%%%%%%%%%%%%%%%%%%%%%%%%%%%%%%%%%%%%%%%%%%
%\markright{\today}
%\pagestyle{myheadings}  
%%%%%%%%%%%%%%%%%%%%%%%%%%%%%%%%%%%%%%%%%%%%%%%%%%%%%%%%%%%%%%%%%%%%
%\def\label#1{\qquad [{\bf\verb@#1@}]}
%\def\ref#1{[\verb@#1@]}
%%%%%%%%%%%%%%%%%%%%%%%%%%%%%%%%%%%%%%%%%%%%%%%%%%%%%%%%%%%%%%%%%%%%
%\def\bibitem#1{\qquad [{\bf\verb@#1@}]}
%\def\cite#1{[\verb@#1@]}
%%%%%%%%%%%%%%%%%%%%%%%%%%%%%%%%%%%%%%%%%%%%%%%%%%%%%%%%%%%%%%%%%%%%
\newcommand{\para}{/\!\!/}
\newcommand{\be}{\begin{equation}}
\newcommand{\ee}{\end{equation}}
\newcommand{\bea}{\begin{eqnarray}}
\newcommand{\eea}{\end{eqnarray}}
\newcommand{\del}{\delta}
\newcommand{\r}{\\ $\rhd$ \hspace{0.5cm}}
\newcommand{\non}{\nonumber}
%%%%%%%%%%%%%%%%%%%%%%%%%%%%%%%%%%%%%%%

%%%%%%%%%%%%%%%%%%%%%%%%%%%%%%%%%%%%%%%
%  Title page
%%%%%%%%%%%%%%%%%%%%%%%%%%%%%%%%%%%%%%%
%\begin{titlepage}
%%%%%%%%%%%%%%%%%%%%%%%%%%%%%%%%%%%%%%%
\def\abstract#1
{\begin{center}{\large Abstract}\end{center}\par #1}
\def\title#1{\begin{center}{#1}\end{center}}
\def\author#1{{\large \begin{center}{#1}\end{center}}}
\def\address#1{\begin{center}{\it #1}\end{center}}
%%%%%%%%%%%%%%%%%%%%%%%%%%%%%%%%%%%%%%%
%%%
\hfill
\parbox{5cm}{{TIT/HEP-\pubnum} \par 
%\today
}
\par 
%%% 
\vspace{15mm}
\title{
{\Large{\bf 
Bubble wall perturbations coupled with gravitational waves 
}} 
}
%%%
\vskip 9mm
\author{
        Akihiro Ishibashi
                \footnote{E-mail:akihiro@th.phys.titech.ac.jp} and
        Hideki Ishihara
                \footnote{E-mail:ishihara@th.phys.titech.ac.jp} 
        }
\address{
        Department of Physics, Tokyo Institute of Technology, \\
        Oh-Okayama Meguro-ku, Tokyo 152, Japan
        }
%%%%%%%%%%%%
\vskip 1.5 cm
\abstract{
We study a coupled system of gravitational waves and a domain wall which is 
the boundary of a vacuum bubble in de Sitter spacetime. 
To treat the system, we use the metric junction formalism of Israel. 
We show that the dynamical degree of the bubble wall is lost and 
the bubble wall can oscillate only while the gravitational waves go across it. 
It means that the gravitational backreaction on the motion of 
the bubble wall can not be ignored. 
}
%%%
\\
PACS number(s) 04.20.-q 04.40.-b 98.80.Cq 98.80.Hw

\maketitle
%\end{titlepage}

%%%%%%%%%%%%%%%%%%%%%%%%%%%%%%%%%%%%%%%%%%%%%%%%%%%%%%%%%%%%
\section{Introduction}

Topological defects could 
be formed when symmetries are broken in the early universe. 
Domain walls, the simplest topological defect, 
emerge at the boundaries of true vacuum bubbles 
nucleating in a false vacuum at first-order phase transitions~\cite{VVS}.  

Motivated by recent observations which suggest that 
the present density parameter $\Omega_0$ is smaller than unity, 
a possible inflationary scenario 
which is compatible with an open universe is proposed~\cite{Open}. 
The scenario requires two inflationary epochs.
First, an old type inflation occurs and solves the horizon 
and the homogeneity problems. 
It is exited through a nucleation of
a single vacuum bubble by quantum tunneling. 
If the tunneling process is represented by the Euclidean 
bounce solution with O(4) symmetry~\cite{CD}, 
the interior of the bubble has the O(3,1) invariance.
This symmetry naturally induces negatively curved time-slices
and an open Friedman-Robertson-Walker (FRW) universe 
is realized inside the bubble. 
Next, a slow rollover inflation occurs after that. 
The present value of $\Omega_0$ is determined by the form 
of the inflaton potential. 
This scenario is called \lq\lq one-bubble open inflation\rq\rq. 
The geometry of the scenario is schematically depicted in Fig.~1.

In the one-bubble open inflation scenario, 
the observable universe is surrounded by a bubble wall. 
A conclusive verification of the scenario is 
the detection of existence of the bubble wall. 
Though the homogeneous isotropic open time-slices do not 
intersect the bubble wall, it is causally 
visible from the observers inside the bubble. 
If the bubble wall is deformed in some reasons, 
we may observe the inhomogeneities of the universe produced by 
the deformation. Thus it is especially relevant to investigate the motion of 
the bubble wall in the one-bubble open inflation context.

In the thin wall approximation, the motion of a domain wall 
is described as a three-dimensional timelike hypersurface 
in the four-dimensional spacetime. 
The small deformation around background domain walls with planar 
or spherical symmetries are represented by a single scalar field 
on the background wall hypersurface~\cite{GVG}. 
The equation for the scalar field takes the form of 
the Klein-Gordon equation on the hypersuface. 
Thus, it is expected intuitively that the wave propagation of the scalar 
field represents the oscillation of the deformed wall. 
These analysis are based on the assumption that all influence of gravity 
on the wall motion is negligible.

In the thin wall approximation, 
the energy-momentum tensor of the wall is concentrated on 
the three-dimensional hypersurface. 
Thus, the spacetime should be singular at the hypersurface. 
In the general relativity, the spacetime containing such a singular 
hypersurface should be treated by the metric junction formalism~\cite{Is}. 
In this method the wall is treated as a source of 
the gravitational field in which the wall moves. 
The motions of domain walls with spherical symmetry 
are analyzed by this formalism in Ref.~\cite{HS,BGG}.  

In the metric junction formalism, Kodama, Ishihara, and Fujiwara studied 
the perturbations of a spherically symmetric domain wall system~\cite{KIF}.
The interaction between the perturbed wall and gravitational waves 
is taken into account in their work. 
As a result, it turned out that the deformation of 
the domain wall occurs only while the gravitational wave goes through 
and does not propagate on the background hypersurface by itself. 

The background configuration of the spacetime considered in
Ref.~\cite{KIF} is constructed in the peculiar way: 
joining two copies of a portion of Minkowski spacetime 
with spherically symmetric boundary by identifying the boundary 
at the domain wall. 
So, there is a reflection symmetry with respect to the spherically symmetric 
domain wall. 
Thus, it is interesting to clarify whether the result is due to 
the symmetric configuration of the background, or to 
the domain wall's feature itself. 
This also motivates us to investigate the motion of domain wall
on the different background configuration: one-bubble open inflation
configuration, in which there exists the vacuum energy deference between 
the inside and the outside of the bubble and no reflection symmetry. 

In this paper, we investigate perturbations of a coupled system 
of gravity and a vacuum bubble in de Sitter spacetime. 
We treat the bubble wall by the thin wall approximation
and apply the metric junction formalism to consider 
the gravitational back reaction on the bubble wall motion. 
We accomplish this in terms of the gauge-invariant perturbation method 
on spherically symmetric background spacetime established by 
Gerlach and Sengupta~\cite{GS1,GS2,GS3}. 
Though, in the main stream of our study, we consider a bubble wall 
whose inside is Minkowski and outside de Sitter spacetime 
as a background, we also examine the case that both sides of the bubble 
are de Sitter spacetimes with different cosmological constants. 
The issue is the extension of that in Ref.~\cite{KIF}.

In the next section, we specify a background geometry which is used in 
the one-bubble open inflation scenario by using metric junction method.
In section 3, we consider the metric perturbations in terms of 
gauge-invariant variables. 
We find master variables for two modes of gravitational waves 
and reduce the perturbed Einstein equations into a single master equation 
for them, and give the general solutions of it. 
In section 4, we write down the perturbed junction conditions as the boundary 
conditions on the master variable. 
We match the general solution along the bubble wall by using the perturbed 
junction conditions and obtain the global solutions.
Section 5 is devoted to summary and discussion.

Throughout this paper, $\kappa$ denotes $8\pi G$ and the unit $c = 1$ is 
taken. 

%%%%%%%%%%%%%%%%%%%%%%%%%%%%%%%%%%%%%%%%%%%%%%%%%%%%%%%%%%%%%%%%%%%%%%%%%%
%
\def\2M{{}^2\!M}
%
%%%%%%%%%%%%%%%%%%%%%%%%%%%%%
\section{Background geometry}
%%%%%%%%%%%%%%%%%%%%%%%%%%%%%

Let us consider a geometrical model in which a spherical vacuum bubble exists
in de Sitter spacetime with a cosmological constant $\Lambda$ 
as a background spacetime. 
The bubble wall divides the spacetime into two regions: 
Minkowski $(M_-, ds^2_-)$ inside and de Sitter spacetime $(M_+,ds^2_+)$ 
outside the bubble. Hereafter, we describe quantities in $M_{-}$ 
with subscript $\lq\lq- "$ and in $M_{+}$, with $\lq\lq + "$.

We assume that the thickness of the bubble wall is
small enough comparing to all other dimensions, that is, 
the thin wall approximation.
Then the history of the bubble wall is described as a three-dimensional
timelike hypersurface $\Sigma$.
The whole spacetime manifold containing the bubble wall is
constructed as $M := M_- \cup \Sigma \cup M_+$
by matching the boundaries
$\Sigma_\pm := \partial M_\pm $ at $\Sigma$. 
The spacetime metric $ds^2$ is required to be $C^0$ in $M$. 
The geometries of the boundaries $\Sigma_\pm$ 
are characterized by the induced intrinsic metrics
$(q_{ij} d\zeta^i d\zeta^j)_\pm $ and the extrinsic curvatures
$(K_{ij} d\zeta^i d\zeta^j)_\pm $, where $\zeta^i_\pm$ 
(the indices $i,j,\cdots$ run over 1,2,3) are three-dimensional
coordinate systems on $\Sigma_\pm $, respectively.
The intrinsic metrics and the extrinsic curvatures are specified 
by the junction conditions.

In more realistic model of the one-bubble open inflationary universe,
the interior of the bubble has a slow rollover inflationary phase 
and it should be treated locally as de Sitter spacetime.
We will see later that the results for this case are essentially the same.
For simplicity, we consider the time symmetric configuration 
as illustrated in Fig.2, instead of a bubble nucleation.  

\subsection{Intrinsic metric and extrinsic curvature}

The background spacetimes have spherical symmetry. 
In general, the metric with spherical symmetry can be expressed in the form 
\be
   ds^2  = g_{\mu\nu}dx^\mu dx^\nu
         = \gamma_{ab}(y^c) dy^a dy^b
           + r^2(y^c)\Omega_{pq}dz^p dz^q,
\label{metric:spherical} 
\ee
where $\Omega_{pq}dz^p dz^q$ is the metric on a unit symmetric two-sphere
with an angular coordinates $z^p$, {\it i.e.}, 
\be
     \Omega_{pq}dz^p dz^q = d\theta^2 + \sin^2\theta d\phi^2
                          =: d\Omega^2 .
\ee
The functions $r(y^c)$ and $\gamma_{ab}(y^c)$
are scalar and tensor fields on the two-dimensional
orbit space $\2M$ spanned by the two-coordinates $y^a$
and each point of $\2M$ represents a symmetric two-sphere.
Hereafter, the indices on the orbit space $a,b,c,\cdots$ run over $0,1$ and 
the indices on the unit two-sphere $p,q,r,\cdots$ over $2, 3$.  

Let $\nabla_\mu$, $D_a$, and $\hat{D}_p$ be the covariant derivatives 
with respect to $g_{\mu \nu}$, $\gamma_{ab}$, and $\Omega_{pq}$, 
respectively. The covariant derivative $\nabla_\mu$ is expressed 
by $D_a$ and $\hat{D}_p$, 
that is, the Christoffel symbol~$\Gamma^{\lambda}_{\mu \nu}$ 
associated with the spacetime metric $g_{\mu \nu}$ is expressed as 
\bea
&&{}
\Gamma^a_{bc} = \tilde{\Gamma}^a_{bc}, \quad
\Gamma^r_{pq} = \hat{\Gamma}^r_{pq}, \nonumber \\
&&{}
\Gamma^p_{aq} = \del^p_q \left(\frac{D_a r}{r}\right) , \quad
\Gamma^a_{pq} = - r^2 \Omega_{pq} \left( \frac{D^a r}{r}\right),
\eea
where $\tilde{\Gamma}^a_{bc}$ and $\hat{\Gamma}^r_{pq}$ are
the Christoffel symbols associated with 
$\gamma_{ab}$ and $\Omega_{pq}$, respectively. 

The orbit space metrics of $M_{\pm}$ can take the static form 
\bea 
  (\gamma_{ab} dy^a dy^b)_{\pm} 
     = - g(r_{\pm}) dt^2_{\pm} + g(r_{\pm})^{-1} dr^2_{\pm}, 
\eea 
where 
\be 
    g(r_{-}) = 1, \qquad g(r_{+}) = 1 - H^2 r^2_{+}, 
\ee 
and $H$ is defined by $H:= \sqrt{\Lambda/3}$. 
The history of the spherically symmetric bubble wall is 
described by a timelike orbit in ${}^2\!M$. 

Let $(n^\mu)_\pm $ be the unit normal vectors to the boundaries
$\Sigma_\pm $, respectively, 
such that both $(n^\mu)_\pm $ direct from $M_-$ to $M_+$. 
Since the vectors $(n^\mu)_\pm $ have vanishing $z^p$ components, 
we can identify them with $(n^a)_\pm$. 
Let $(\tau^{a})_\pm $ be the future-directed unit timelike
tangential vectors to $\Sigma_\pm $ on the orbit space. 
Then the orbit space metric can be decomposed as
\be
        (\gamma_{ab})_\pm  
                = - (\tau_a)_\pm  (\tau_b)_\pm + (n_a)_\pm (n_b)_\pm .
\ee 

The projection tensors $(q_{\mu\nu})_\pm$ onto $\Sigma_\pm$ are represented by 
\bea
    (q_{\mu \nu})_\pm
              &:=& (g_{\mu \nu})_\pm - (n_\mu)_\pm (n_\nu)_\pm 
\non \\
              &=& - (\tau_\mu)_\pm (\tau_\nu)_\pm + r^2_\pm \Omega_{\mu \nu},
\label{eq:proj}
\eea
where $\tau^\mu$ and $\Omega_{\mu \nu}$ are the four-dimensional extensions 
of $\tau^a$ and $\Omega_{pq}$. 
The four-dimensional extensions of the extrinsic curvatures 
$(K_{\mu \nu})_\pm $ are defined as
\bea
        (K_{\mu \nu})_\pm
                &:=& - (q_\mu^\lambda q_\nu^\sigma
                                        \nabla_\lambda n_\sigma)_\pm. 
\label{eq:extc} 
\eea 
From Eqs.~(\ref{eq:proj}) and (\ref{eq:extc}), 
the extrinsic curvatures $(K_{\mu \nu})_\pm $ can be represented by
\bea
        (K_{ab})_\pm &=& - (\tau_a\tau_b \tau^c D_{\para}n_c)_\pm, \\
        (K_{ap})_\pm &=& 0, \\
        (K_{pq})_\pm &=& - (r {D_{\perp} r})_\pm  \Omega_{pq},
\label{Kpq}
\eea
where 
\be
        D_{\para} := \tau^a D_a,\quad D_{\perp} := n^a D_a. 
\label{eq:Dp}
\ee  

\subsection{Background bubble motion}

By using the metric junction formalism~\cite{Is}, 
the motions of spherically symmetric thin walls are studied in Ref.~\cite{HS}. 
To establish notation, we briefly review the metric junction formalism 
and determine the motion of a spherical vacuum bubble in de Sitter spacetime 
following Ref.~\cite{BGG}. 

For convenience, we introduce the Gaussian normal coordinate system 
$(x^\mu) = (\chi,\zeta^i)$ in the neighborhood of the hypersurface $\Sigma$ 
such that the spacetime metric takes the form
\begin{equation}
        ds^2 = g_{\mu \nu} dx^{\mu}dx^{\nu} 
                        = d\chi^2 + q_{ij}d\zeta^i d\zeta^j. 
\label{GNC}
\end{equation}
The coordinate axis ${\chi}$ is normal to $\Sigma$ and 
can be set $\chi = 0 $ on $\Sigma$.

Let $S_{\mu \nu}(\zeta^i)$ be the wall stress-energy tensor which is defined 
by the energy-momentum tensor $T_{\mu \nu}$ as 
\be  
  S_{\mu \nu}(\zeta^i) 
       :=\lim_{\epsilon \to 0 } \int^\epsilon_{-\epsilon} T_{\mu \nu} d\chi 
\label{def:wset}
\ee
in the thin wall approximation.
Then the energy-momentum tensor can be described as
\be
   T_{\mu \nu} 
        = S_{\mu \nu}(\zeta^i) \delta (\chi) - \rho g_{\mu \nu} \theta(\chi), 
\label{emtensor}
\ee
where $\rho = 3 H^2/\kappa $ is the vacuum energy of de Sitter spacetime 
and $\theta(\chi)$ is a step function.
Provided that the spacetime metric $g_{\mu \nu}$ is continuous at $\chi = 0$, 
$(q_{ij})_+$ coincides with $(q_{ij})_-$ on $\Sigma$. 
Hence, 
\be
        \left[q_{ij} \right] = 0, 
\label{continu}
\ee
where, for geometrical quantity $Q$ on $M_\pm$, $\left[Q \right]$ denotes
\be
  \left[Q \right] 
    := \lim_{\epsilon \to 0}\{ Q|_{\chi = +\epsilon} - Q|_{\chi = -\epsilon}\}.
\ee

With the help of the Gauss-Codazzi equation,
the $(i,j)$ components of the Einstein equations are written by
\begin{equation}
%{}^{(4)}\!R_{ij} = 
{}^{(3)}\!R_{ij} - KK_{ij} + 2K_{il}K^l_j + \partial _{\chi}K_{ij}
        = \kappa (T_{ij} - \frac{1}{2} g_{ij} T), 
\label{eq:GC}
\end{equation}
where $^{(3)}\!R_{ij}$ is the Ricci curvature on $\Sigma$ with respect to 
$q_{ij}$. 
Substituting Eq.~(\ref{emtensor}) into this equation~(\ref{eq:GC}) 
and integrating it with respect to $\chi$ in an infinitesimally small 
interval $(- \epsilon, + \epsilon)$, we obtain 
\be
\left[ K^i_j \right] = \kappa \left( S^i_j
- \frac{1}{2} \del^i_j S \right). \label{jc:Kij}
\ee

Similarly, from the $(\chi, i)$ and the $(\chi,\chi)$ components, we obtain
\bea
{}&&
\overline{K}_{ij}S^{ij} = \rho, \label{jc:Trace}\\
{}&&
{}^{(3)}D_j S^{ij} = 0, \label{jc:Div}
\eea
where, for $Q$ on $M_\pm$, $\overline{Q}$ denotes the mean value 
\be
\overline{Q} := \frac{1}{2}
\lim_{\epsilon \to 0} \{Q|_{\chi = + \epsilon}
+ Q|_{\chi = - \epsilon} \},
\ee
and ${}^{(3)}\!D_i$ is the covariant derivative with respect to $q_{ij}$.

Assuming that a scalar field comprises the bubble wall, 
we can write the wall stress-energy tensor $S_{\mu \nu}$ by a single scalar 
field $\sigma$~\cite{BGG} as
\be
        S_{\mu \nu} = - \sigma q_{\mu \nu}. 
\label{eq:stress}
\ee
From the condition~(\ref{jc:Div}), $\sigma$ should be constant. 
Then, the junction conditions~(\ref{jc:Kij}) and (\ref{jc:Trace}) reduce to
\be
        \left[ K^i_j \right] = \frac{1}{2} \kappa \sigma \delta ^i_j  , 
\label{eq:JCKij}
\ee
and
\be
        \overline{K} = - \frac{\rho}{\sigma}. 
%                    = - \frac{3H^2}{\kappa \sigma}.
\label{eq:JCK}
\ee
The bubble wall motion is determined by the conditions~(\ref{continu}) 
and (\ref{eq:JCKij}). Eq.~(\ref{eq:JCK}) is then satisfied automatically. 

The unit tangential vectors to $\Sigma_\pm$ are expressed as
\be
        \tau^a_\pm := dy^a_\pm / d{\tau}
                = (\dot{t} ,\dot{r})_\pm  
\ee
in the static charts $(t, r)_\pm$ of ${}^2\!M$. 
Here, ${\tau}$ is a propertime along the wall orbit and a dot 
denotes the derivative with respect to ${\tau}$.
From the normalization $ \tau^a \tau_a = -1 $, it follows that 
\bea
        \dot{t}_{\pm} = \epsilon \frac{1}{g(r_{\pm})}
                \sqrt{g(r_{\pm}) + \dot{r}^2_{\pm}}, \quad (\epsilon=\pm 1) .  
\eea
From the orthogonality  $n_a \tau^a= 0$ and the normalization $n_a n^a = 1$, 
the components $(n^t, n^r)_\pm$ are written as 
\bea
        (n^t)_{\pm}     = \frac{\dot{r}_{\pm}}{g(r_{\pm})}, \qquad 
        (n^r)_{\pm}     = \epsilon \sqrt{ g(r_{\pm}) + \dot{r}^2_{\pm}}
                = g(r_{\pm}) \dot{t}_{\pm}  . 
\eea
Then, from Eq.~(\ref{Kpq}), we get the $(p,q)$ components 
of the extrinsic curvatures of $\Sigma_{\pm}$ in the form 
\bea 
  &&(K^p_q)_{\pm} 
  = - \epsilon \frac{\sqrt{g(r_{\pm}) + \dot{r}^2_{\pm}}}{r_{\pm}}\del^p_q. 
\label{Kpq+-}
\eea

The $(p,q)$ components of the continuity condition~(\ref{continu}) require 
$r_+({\tau}) = r_-({\tau}) = r({\tau})$. 
Then, substituting Eq.~(\ref{Kpq+-}) into the condition~(\ref{eq:JCKij}), 
we obtain 
\be
        1 + \dot{r}^2 = \left(\frac{r}{\alpha}\right)^2, 
\label{eq:r}
\ee
where $\alpha$ is the constant defined by 
\be
        \alpha := \frac{4 \kappa \sigma}{4H^2 + \kappa^2 \sigma^2}.  
\label{def:alpha}
\ee
The value of the constant $\alpha$ gives the minimal radius of the bubble, 
which is the radius on the instance of the bubble nucleation. 

The solution of Eq.~(\ref{eq:r}) is obtained as
\be
        r = \alpha \cosh{\tau}. 
\ee
Here and for the rest of this paper, $\tau$ denotes the normalized 
propertime by $\alpha $.
Then from Eq.~(\ref{eq:proj}), the intrinsic metric 
on the wall takes the form
\be
   q_{ij}d\zeta^i d\zeta^j = \alpha^2 (- d\tau^2 + \cosh^2 \tau d\Omega^2).
\label{eq:dS3}
\ee
This is the three-dimensional de Sitter spacetime metric 
with the O(3,1) symmetry of the bubble wall.

As a natural extension of the wall metric~(\ref{eq:dS3}), we introduce 
the hyperbolic charts $(\tau_\pm ,\chi_\pm ,\theta_\pm ,\phi_\pm )$ defined by 
\bea 
  && t_-  =: a(\chi_- ) \sinh \tau_- , \quad 
     \sqrt{g(r_{+})} \sinh Ht_+ =: a(\chi_+ ) \sinh \tau_+ , \quad \non \\ 
  && r_\pm  =: a(\chi_\pm ) \cosh \tau_\pm , 
\label{eq:Hyperchart} 
\eea 
where
\begin{equation}
 a(\chi_\pm ) := \left\{
           \begin{array}{@{\,}ll}
               (1/H)\sin H\!{\chi_+}
                        & \mbox{for}
                \quad \chi > 0 \mbox{ --- de Sitter side,} \\
               \chi_- & \mbox{for}
                \quad \chi < 0 \mbox{ --- Minkowski side.}  
           \end{array}
        \right. 
\label{scale}
\end{equation}
This is a Gaussian normal coordinate system~(\ref{GNC}) of the considering 
model. The coordinates $\chi_\pm $ are related to $\chi$ as
\be
\chi = a(\chi_\pm ) - \alpha,
\ee
and hence at the bubble wall
\be
a(\chi_\pm )|_{\Sigma} = \alpha.
\ee
In addition, on the wall, it can be set naturally as
\begin{equation}
\tau_\pm  = \tau, \quad \theta_\pm  = \theta, \quad
\phi_\pm  = \phi.
\end{equation} 
Then, in the hyperbolic charts, the spacetime metrics take the form 
\be
(ds^2)_\pm = d\chi_\pm ^2 + a^2(\chi_\pm ) 
                  \left\{ 
                         - d\tau^2 + \cosh^2 \tau d\Omega^2 
                  \right\}.  
\label{Charts:Hyp}
\ee  

Since these coordinate systems, which are also called 
the spherical Rindler charts, do not cover the whole spacetime, 
an analytic extension is necessary to cover the interior of the forward
light cone from the center of the bubble, 
which corresponds to the observable universe. 
By the coordinate transformations,
\be
\chi_- = i\eta, \quad \tau_- = \xi - i \frac{\pi}{2},
\label{ext:chart}
\ee
the metric takes the form
\be
ds^2_- = - d\eta^2 + \eta^2 \left\{ 
                             d\xi^2 + \sinh ^2 \xi d\Omega^2    
                      \right\}.
\ee
This is the Milne universe, which has open FRW time-slices. 

In the hyperbolic charts, the extensions of the normal vectors
$(n^{\mu})_\pm $ and the tangential vectors $(\tau^{\mu})_\pm $
to $\chi = \mbox{const.}$ hypersurfaces in a neighborhood
of $\Sigma$ are expressed, respectively, as
\be
(n^{\mu})_\pm  = (\partial_{\chi_\pm })^{\mu}, \qquad
(\tau^{\mu})_\pm  = \frac{1}{a(\chi_\pm )}(\partial_{\tau})^{\mu}.
\ee
The extrinsic curvatures of the $\chi = \mbox{const.}$ hypersurfaces 
are calculated explicitly as
\begin{equation}
(K_{ij})_\pm  = - ( \partial_{\chi_\pm }\ln a(\chi_\pm ) )(q_{ij})_\pm, 
\label{eq:extcurva}
\end{equation} 
and evaluated on $\Sigma_{\pm}$ as
\bea
%&&{}
(K_{ij})_- = - \frac{1}{\alpha} (q_{ij})_-, \qquad
%&&{}
(K_{ij})_+ = - \frac{\sqrt{1 - H^2 \alpha^2}}{\alpha} (q_{ij})_+. 
\label{eq:K}
\eea

%%%%%%%%%%%%%%%%%%%%%%%%%%%%%%%%%%%%%%%%%%%%%%%%%%%%%%%%%%%%%%%%%%%%%%%%%%%%
\section{Perturbations}

In this section, we consider the metric perturbations on the background
geometry constructed in the previous section. We describe the metric 
perturbations by the tensor harmonics defined in Appendix A. 
We introduce gauge-invariant variables in accordance with
Gerlach and Sengupta~\cite{GS1} and express the perturbed Einstein equations
in terms of them. Furthermore, we show that the perturbed Einstein equations 
reduce to a single scalar master equation in ${}^2\!M$.   

\subsection{Gauge-invariant perturbation variables}
Let us consider the metric perturbations 
$(h_{\mu \nu}dx^{\mu}dx^{\nu})_{\pm}$ on the background metrics 
$(g_{\mu \nu}dx^{\mu}dx^{\nu})_{\pm}$. 
Due to the spherical symmetry of the background geometry, 
we can expand the perturbations $(h_{\mu \nu}dx^{\mu}dx^{\nu})_{\pm}$
by using the tensor harmonics on the unit sphere (see Appendix A, for 
their definitions) as follows: 
\bea 
\mbox{for odd modes,} 
&&{}
\non \\
h^{(o)}_{\mu \nu}dx^{\mu}dx^{\nu}
  &=& 
     \sum_{l,m} 
     \left\{
         r (f^{(o1)}{}_{lm})_{a}(V^{(o)}{}_{lm})_{p} 
                                   \left( dy^{a}dz^{p} + dz^{p}dy^{a} \right) 
        + r^2 (f^{(o2)}{}_{lm})(T^{(o2)}{}_{lm})_{pq} 
                                               dz^{p}dz^{q}
     \right\}, 
\label{oddharmoexp}
\\ 
\mbox{for even modes,} 
&&{} 
\non \\ 
h^{(e)}_{\mu \nu}dx^{\mu}dx^{\nu}
  &=&
    \sum_{l,m} 
    \Bigl\{
          (f^{(e)}{}_{lm})_{ab}Y_{lm} 
                                   dy^{a}dy^{b} 
          + r(f^{(e1)}{}_{lm})_{a}(V^{(e)}{}_{lm})_{p}
                                  \left( dy^{a}dz^{p} + dz^{p}dy^{a} \right) 
\non \\
& & {} + \left( 
              r^2 (f^{(e0)}{}_{lm}) (T^{(e0)}{}_{lm})_{pq} 
              + r^2 (f^{(e2)}{}_{lm}) (T^{(e2)}{}_{lm})_{pq} 
         \right) 
               dz^{p}dz^{q}
    \Bigr\}. 
\label{evenharmoexp}
\eea 
The subscripts $e$ and $o$ denote the even and the odd modes,
respectively.
The expansion coefficients, $f(y^c), f_a(y^c), f_{ab}(y^c)$ are
scalar, vector, and symmetric tensor fields on the orbit space ${}^2\!M$,
respectively. 
Hereafter we suppress the angular integers $l,m$ and summation $\sum_{l,m}$ 
for notational simplicity and discuss the modes $l \ge2$, 
which are relevant to the gravitational waves.

The metric perturbations $h_{\mu \nu}dx^{\mu}dx^{\nu}$ have the gauge freedoms.
Associated with an infinitesimal coordinate transformations, 
$x^{\mu} \rightarrow  x^{\mu} + \xi^{\mu}$, 
the gauge transformed metric perturbations are
\bea 
\bar{h}_{\mu \nu}dx^{\mu}dx^{\nu} 
&=&{} 
     h_{\mu \nu}dx^{\mu}dx^{\nu} + \bar\del h_{\mu \nu}dx^{\mu}dx^{\nu}, 
\non \\  
&=&{}
     h_{\mu \nu}dx^{\mu}dx^{\nu} - \left( \nabla_{\mu} \xi_{\nu}
                + \nabla_{\nu} \xi_{\mu}\right) dx^{\mu}dx^{\nu}. 
\label{gaugeprtrb}
\eea 
The generator $\xi_{\mu}dx^{\mu}$ of the infinitesimal coordinate 
transformations is expanded by the tensor harmonics as
\bea
&&{}
\xi^{(o)}_{\mu}dx^{\mu} = r \xi^{(o)} V^{(o)}_{p} dz^{p}, \\
&&{}
\xi^{(e)}_{\mu}dx^{\mu} = \xi^{(e)}_{a} Y dy^{a}
+ r \xi^{(e)} V^{(e)}_p dz^{p}.
\eea
Then, from Eqs.~(\ref{oddharmoexp}), (\ref{evenharmoexp}), 
and (\ref{gaugeprtrb}), the gauge transformed expansion coefficients 
are expressed as follows: 
\bea 
\mbox{for odd modes,}
&&{} 
\non \\
&&{}
    \bar{\del} f^{(o1)}_{a}  
                = - r D_a \left( \frac{\xi^{(o)}}{r} \right),
\label{o1}\\
&&{}
    \bar{\del} f^{(o2)} 
                = - \frac{2}{r} \xi^{(o)},  
\label{o2} 
\\ 
\mbox{for even modes,} 
&&{}
\non \\ 
&&{}
\bar{\del} f^{(e)}_{ab} = - D_{a} \xi_{b} - D_{b} \xi_{a}, \label{e1} \\
&&{}
\bar{\del} f^{(e1)}_{a} = - r D_a \left( \frac{\xi^{(e)}}{r} \right)
- \frac{1}{r}\xi_a, \label{e2} \\
&&{}
\bar{\del} f^{(e0)} = 2 \frac{l(l+1)}{r}\xi^{(e)}
- \frac{4}{r} \xi^{a} D_{a}r,  \label{e3} \\
&&{}
\bar{\del} f^{(e2)} = - \frac{2}{r} \xi^{(e)}. \label{e4}
\eea

According to Gerlach and Sengupta~\cite{GS1}, 
let us introduce the gauge-invariant variables by combining
the expansion coefficients of the metric perturbations.  

From Eqs.~(\ref{o1}) and (\ref{o2}), we take the odd mode gauge-invariant
perturbation variables 
\begin{equation}
{\cal F}_a 
         := f^{(o1)}_{a} - \frac{1}{2} r D_a f^{(o2)}. 
\label{def:Fa} 
\end{equation}
  
From Eqs.~(\ref{e1})--(\ref{e4}), 
we take the even mode gauge-invariant variables
\be
{\cal F}_{ab} 
      :=  f_{ab} - D_a X_b - D_b X_a + \frac{1}{2} \gamma_{ab} 
      \left\{ f^{(e0)} + l(l+1) f^{(e2)} - \frac{4}{r} X^c D_c r \right\}, 
\label{def:Fab} 
\ee
where the vector $X^{a}$ is defined by
\begin{equation}
X^{a} := r f^{(e1)a} - \frac{1}{2} r^2 D^{a} f^{(e2)}. 
\label{def:Xa} 
\end{equation}
From Eqs.~(\ref{e2}) and (\ref{e4}), we see that 
the vector $X^{a}$ transforms as
\be
   \bar{\del} X^{a} = -\xi^{a}. 
\label{eq:X}
\ee

\subsection{The perturbed Einstein equations}
The perturbed Einstein equations on de Sitter background spacetime 
are described by
\bea
&&{} 2 \del G_{\mu \nu} \equiv
- \Box h_{\mu \nu} - \nabla_{\mu}\nabla_{\nu} h^{\lambda}_{\lambda}
+ g_{\mu \nu} \Box h^{\lambda}_{\lambda}
+ \nabla_{\mu} \nabla_{\lambda} h^{\lambda}_{\nu}
+ \nabla_{\nu} \nabla_{\lambda} h^{\lambda}_{\mu} \nonumber \\
&&{}
\qquad \qquad
- g_{\mu \nu} \nabla_{\lambda} \nabla_{\sigma} h^{\lambda \sigma}
+ H^2 g_{\mu \nu}h^{\lambda}_{\lambda} - 4H^2 h_{\mu \nu} = 0, 
\label{eq:pEe}
\eea
where $\Box := \nabla^{\mu} \nabla_{\mu}$.
We express this perturbed Einstein equations by the gauge-invariant variables 
introduced in Eqs.~(\ref{def:Fa}) and (\ref{def:Fab}). 
Further, we rewrite them in the form of the wave equation 
for a single scalar field on the orbit space.

For a while, we consider only the de Sitter side $(M_+, ds^2_+)$ and 
suppress the suffix $\lq\lq + "$. 
All geometrical quantities for Minkowski side $(M_-, ds^2_-)$ are 
reproduced by taking the limit $H \rightarrow 0$. 
We shall discuss the odd and the even modes separately.  

\subsubsection{Odd modes}
The odd mode perturbed Einstein equations are written by the odd mode 
gauge-invariant variables in the form
\be
D_c \left\{ 
            r^4 D^c \left( \frac{{\cal F}^a}{r}\right)
            - r^4 D^a \left( \frac{{\cal F}^c}{r}\right) 
    \right\} 
            - (l-1)(l+2) r{\cal F}^a = 0, 
\label{eq:rotFo}
\ee
\be
  D^a(r{\cal F}_a) =0. 
\label{eq:divFo}
\ee 

From Eq.~(\ref{eq:divFo}), we can find a scalar variable
$\Phi_{(o)}$ such that
\be
r{\cal F}^a = \epsilon^{ab}D_{b} \Phi_{(o)}, 
\label{eq:oPhi}
\ee  
where $\epsilon^{ab} := n^a \tau^b - \tau^a n^b $ is the antisymmetric 
two-tensor. 
Substituting Eq.~(\ref{eq:oPhi}) into Eq.~(\ref{eq:rotFo}), 
we obtain 
\be
\left\{
r^2 \Box_{2} - 2 r (D^{c}r) D_{c} - (l-1)(l+2) \right\} \Phi_{(o)} = 0,
\ee
where $\Box_{2}:= D^{a}D_{a}$.
We can further rewrite this equation in the form 
\be
\left( 
      - \Box_{2} + \frac{l(l+1)}{r^2} 
\right)
       \left( 
             \frac{\Phi_{(o)}}{r} 
       \right) = 0.  
\label{eq:omaster}
\ee  
This is the odd mode master equation for $\Phi_{(o)}$. 

\subsubsection{Even modes}
The even mode perturbed Einstein equations are written by the even mode
gauge-invariant variables as the following set of equations: 
\bea
&&{} 
    D_b {\cal F}^b_a = D_a {\cal F}_b^b, 
\label{Ein:constraint} 
\\
&&{} 
\frac{1}{2}
  \left\{ \Box_{2} + 2\left(\frac{D^c r}{r}\right) D_c 
          - \frac{(l-1)(l+2)}{r^2} 
  \right\} {\cal F}_{ab} 
       -  \left( \frac{D^c r}{r} \right)\left( \frac{D^d r}{r} \right)
                                             {\cal F}_{cd} \gamma_{ab} 
       -  \left( \frac{D^c r}{r} \right)\left( \frac{D_c r}{r} \right)
                                             {\cal F}_{ab}  
\non \\ 
&&{} 
\qquad \qquad \qquad \qquad 
       +  \left( \frac{D^c r}{r} \right)\left( \frac{D_c r}{r} \right)
                                              {\cal F}^d_d \gamma_{ab} 
       -  \left( \frac{D^c r}{r} \right) 
                  \left( D_b {\cal F}_{ac} + D_a {\cal F}_{bc} \right) 
\non \\
&&{}
\qquad \qquad \qquad \qquad 
       +  \left( \frac{D_a r}{r} \right) D_c {\cal F}^c_b 
       +  \left( \frac{D_b r}{r} \right) D_c {\cal F}^c_a 
       = 0. 
\label{Ein:evolve}
\eea  
Eq.~(\ref{Ein:constraint}) implies that there exists a master
scalar variable $\Phi_{(e)}$ such that
\bea 
&&{}
{\cal F}_{ab} 
  =  \left( 
            D_a D_b + H^2 \gamma_{ab} 
     \right) \Phi_{(e)}. 
\label{exist:pote} 
\eea 
Substituting Eq.~(\ref{exist:pote}) into Eq.~(\ref{Ein:evolve}), 
we obtain 
\bea 
  \left( 
        - \Box_{2} + \frac{l(l+1)}{r^2} 
  \right)
         \left( 
               \frac{\Phi_{(e)}}{r} 
         \right) = 0. 
\label{eq:emaster}
\eea 
For the derivation of Eq.~(\ref{eq:emaster}), see Appendix~B. 
This is the even mode master equation 
and is the same form with the odd mode master equation~(\ref{eq:omaster}). 

\subsection{General solutions of the master equation}
Let us give general solutions of the odd and the even mode master 
equations. 
Here, we omit the subscripts $e$ and $o$, since the both 
master equations~(\ref{eq:omaster}) and (\ref{eq:emaster}) take the same form.
We express the master equations in the hyperbolic charts 
defined by Eq.~(\ref{eq:Hyperchart}) in the form  
\be
\left( - a(\chi_{\pm})^2 \partial^2_{\chi_{\pm}} 
      - a( \chi_{\pm}) \left( \partial_{\chi} a( \chi_{\pm}) \right) 
      \partial_{\chi_{\pm}} + \partial^2_{\tau} + \frac{l(l+1)}{\cosh^2\tau} 
\right)
       \left( \frac{\Phi}{r}\right)_{\pm} = 0. 
\label{eq:Mater}
\ee
Assuming that $(\Phi/r)_{\pm}$ can be decomposed
into $R(\chi_{\pm}) T(\tau)$, we obtain the equations for 
$R(\chi_{\pm})$ and $T(\tau)$, respectively,
\bea 
&&{}
\left(
      \frac{1}{a(\chi_{\pm})}\partial_{\chi_{\pm}} a(\chi_{\pm})
      \partial_{\chi_{\pm}} + \frac{k_{\pm}^2}{a(\chi_{\pm})^2}
\right) R(\chi_\pm) = 0, 
\label{eq:MasterR}
\\
&&{}
\left(
      \partial^2_{\tau} + \frac{l(l+1)}{\cosh^2 \tau} + k_{\pm}^2
\right) T(\tau) = 0, 
\label{eq:MasterT} 
\eea 
where $k_{\pm}$ are the separation constants. 
The solutions of Eq.~(\ref{eq:MasterR}) are given by
\be
R(\chi_{\pm}) = e^{\pm ik_{\pm} \zeta(\chi_{\pm})}, 
\label{Rchi}
\ee
where
\bea
&&{}
\zeta(\chi_{-}) := \ln {\chi_{-}}, \\
&&{}
\zeta(\chi_{+}) := - \ln \left\{ \tan \left(\frac{H\chi_{+}}{2}\right)\right\}.
\eea 

The solutions of Eq.~(\ref{eq:MasterT}) are given by
\bea
T(\tau) &\propto& P^{\pm ik_{\pm}}_{l}(-\tanh \tau) \label{Legendre} \\
        &\propto& G_{l} \left( 1, 1\mp ik_{\pm}; \frac{1+\tanh \tau}{2}
\right) e^{\mp ik_{\pm} \tau}, \label{Jacobi}
\eea
where $P_l$ and $G_l$ are the Legendre function and the Jacobi polynomial.

Let us introduce the set of functions $\{\psi_{l}(\tau, k)\}$ defined by 
\begin{equation}
\psi_{l}(\tau,k) := G_{l} \left( 1, 1\mp ik; \frac{1+\tanh \tau}{2} \right)
 e^{- ik \tau}.
\end{equation}
This satisfies the orthogonality condition
\be
\int^{\infty}_{-\infty} d\tau \psi^*_{l}(\tau,k')
               \psi_{l}(\tau,k) = 2 \pi \delta(k - k'), 
\label{eq:ortho}
\ee  
and the set $\{\psi_{l}(\tau, k)\}\,(-\infty < k < \infty)$ is complete 
in the $L^2(- \infty, \infty)$ space of functions of $\tau$. 
Then we can express the solutions $(\Phi /r)_{\pm}$ as
\be
\left(\frac{\Phi}{r}\right)_{\pm}
      = \int^{\infty}_{-\infty} dk \psi_{l}(\tau,k)
\left\{
        A_{\pm}(k) e^{ik\zeta(\chi_{\pm})}
        + B_{\pm}(k) e^{-ik \zeta(\chi_{\pm})}
\right\}. 
\label{eq:sol}
\ee
The mode expansion coefficients $A_{\pm}$ represent the gravitational waves 
propagating from the past Rindler horizon $\tau = - \infty$ toward the bubble
wall and $B_{\pm}$ the waves propagating from the bubble wall toward 
the future Rindler horizon $ \tau = \infty$, as depicted schematically
in Fig.~3.

\section{Global Solutions}
In this section, we first write down the perturbed junction conditions.
We can express them in terms of the master variables. 
Then, by matching the solutions of the master equations 
in $M_{-}$ and $M_{+}$ along the bubble wall, 
we construct the global solutions of the perturbed Einstein equations 
in the whole spacetime $M$. 

The solutions of the perturbed Einstein equations~(\ref{eq:pEe}) should 
satisfy the following perturbed junction conditions 
\bea
&&{}
[\delta q_{ij}] = 0, \label{eq:delqij} \\
&&{}
[\delta K^i_j ] = \kappa (\delta S^i_j - \frac{1}{2} \delta^i_j \delta S).
\eea
Since the wall stress-energy tensor $S_{\mu \nu}$ is given by 
Eq.~(\ref{eq:stress}) with the constant $\sigma$, 
the second condition reduces to
\be
    \left[ 
           \delta K^i_j 
    \right] = 0. 
\label{eq:delKij}
\ee

As in the case of the metric perturbations, we can expand the perturbed
extrinsic curvature $\del K_{ij}$ by the tensor harmonics:
\bea
\mbox{for odd modes,}
&&{} 
\non 
\\
&&{} 
(\del K^{(o)})_{ij}dx^idx^j = r (\del K^{(o1)})_{a} (V^{(o)})_{p}
( dy^a dz^p + dz^p dy^a) + r^2 (\del K^{(o2)}) (T^{(o2)})_{pq}dz^p dz^q, 
\\
\mbox{for even modes,}
&&{} 
\non 
\\
&&{}
(\del K^{(e)})_{ij}dx^idx^j = (\del K^{(e)})_{ab}Y dy^a dy^b +
r (\del K^{(e1)})_{a} (V^{(e)})_{p} ( dy^a dz^p + dz^p dy^a) \nonumber \\
&&{}
\qquad \qquad \qquad \qquad
+ r^2 \left\{ (\del K^{(e0)})(T^{(e0)})_{pq}  + (\del K^{(e2)}) (T^{(e2)})_{pq}
\right\} dz^p dz^q.
\eea

For convenience, let us define the decompositions of tensors on ${}^2\!M$ 
by using $\tau^a$ and $n^a$. A vector $V^a$ is decomposed as
\begin{equation}
V^a = - V_{\para}\tau^a + V_{\perp} n^a,
\end{equation}
where
\begin{equation}
V_{\para}:= \tau^a V_a, \quad V_{\perp} := n^a V_a .
\end{equation}
Similarly, a symmetric tensor $T_{ab}$ is decomposed as
\begin{equation}
T_{ab} = T_{\para \para}\tau_a \tau_b + T_{\perp \perp}n_a n_b
  - T_{\para \perp}(\tau_a n_b + n_a \tau_b),
\end{equation}
where
\be
T_{\para \para} := \tau^a \tau^b T_{ab}, \quad
T_{\perp \perp} := n^a n^b T_{ab}, \quad
T_{\para \perp} := \tau^a n^b T_{ab}.
\ee 
We also define the projections of the perturbed extrinsic curvature
\be
(\delta K_{\para})^{\mu}_{\nu} := q^{\mu}_{\lambda}\delta K^{\lambda}_{\nu},
\label{Proj:Kmunu}
\ee
which are continuous from Eq.~(\ref{eq:delKij}). 

Now let us express Eq.~(\ref{Proj:Kmunu}) in terms of the gauge-invariant 
variables as follows~\cite{GS2,GS3}: 
\bea
\mbox{for odd modes,}
&&{} 
\non \\
&&{} 
(\del K^{(o1)}_{\para})^{a} \tau_{a} 
   = \frac{1}{2} \epsilon^{ab}
            D_{b}\left(\frac{{\cal F}_{a}}{r} \right), 
\label{eq:odd1}\\
&&{}
  (\del K^{(o2)}_{\para}) = \frac{{\cal F}_{\perp}}{r}. 
\label{eq:odd2}
\\ 
\mbox{for even modes,} 
\non \\ 
&&{}
 (\del K^{(e)}_{\para})^a_a
         = - D_{\para} {\cal F}_{\para \perp} 
           + \frac{1}{2} D_{\perp} \left( {\cal F}_{\perp \perp} 
           - \frac{1}{2}{\cal F}^c_c \right) 
           - \frac{1}{2}(K^a_a)
                    \left( 
                          {\cal F}_{\perp \perp}  - \frac{1}{2}{\cal F}^c_c 
                    \right) 
\non \\ 
&&{}{\qquad \qquad \qquad} 
          + \{- D^2_{\para} + (K^a_a)^2 + H^2 \}X_{\perp},
\label{eq:Kaa1}
\\
&&{}
(\del K_{\para}^{(e1)})^a \tau_a
         = \frac{1}{2} {\cal F}_{\para \perp} 
           + r D_{\para} \left( \frac{X_{\perp}}{r} \right),
\label{eq:Kaa2}\\
&&{}
(\del K_{\para}^{(e0)}) 
         = 
           - \frac{1}{2}(K^p_p) 
           \left( 
                 {\cal F}_{\perp \perp} - \frac{1}{2}{\cal F}^c_c 
           \right) 
           - 2 \left( \frac{D_{\para}r}{r}\right) {\cal F}_{\para \perp} 
           - \frac{1}{2}D_{\perp} {\cal F}^c_c 
\nonumber \\
&&{}
\qquad \quad 
       +\left\{
               -2 \left(\frac{D_{\para}r}{r}\right) 
                             D_{\para} - \frac{l(l+1)}{r^2}
                             + \frac{1}{2}(K^p_p)^2 + 2 H^2
        \right\} X_{\perp}, \label{eq:Kaa3}  \\
&&{}
(\del K_{\para}^{(e2)}) = \frac{X_{\perp}}{r^2}. 
\label{eq:Kaa4}
\eea

We note that these quantities contain the gauge dependent 
variable~$X_{\perp}$. 
In the perturbation theory, we treat the quantities on the perturbed geometry
by embedding to the background geometry. 
In the present case, we treat the geometry constructed by gluing two 
different spacetimes: Minkowski spacetime and de Sitter spacetime. 
In addition, the geometry contains a singular hypersurface (the bubble wall) 
at the boundaries of the two spacetimes. 
We therefore must choose the embedding such that it maps the perturbed 
geometry of Minkowski (de Sitter) spacetime to the background 
Minkowski (de Sitter) spacetime and furthermore the perturbed bubble wall 
to that on the background. 
Thus, we consider the gauge transformations under the restriction 
such as $\xi_{\perp} = 0$ on the bubble wall. 
The variable $X_{\perp}$ is gauge-invariant just for the gauge 
transformations on the bubble wall and can be interpreted as 
the wall displacement variable~\cite{KIF}. 
This is discussed in a separate work~\cite{IsIs}. 

%
%Using Eqs.~(\ref{eq:odd1}), (\ref{eq:odd2}), (\ref{eq:Kaa1}), 
%(\ref{eq:Kaa2}), (\ref{eq:Kaa3}), and (\ref{eq:Kaa4}), 
%
Using Eqs.~(\ref{eq:odd1})--(\ref{eq:Kaa4}), 
we can rewrite the junction condition~(\ref{eq:delKij}) 
in terms of the gauge-invariant variables and
further in terms of the master variables $\Phi_{(o)}$ and $\Phi_{(e)}$. 
We consider the odd and the even modes separately 
in the following subsections. 

\subsection{Odd modes} 
First, from Eq.~(\ref{eq:delqij}), we see that $f^{(o2)}$ and
$f^{(o1)}_{\para }$ are continuous. Then, 
\be
\left[ {\cal F}_{\para} \right] 
     = \left[ f_{\para}{}^{(o1)} - \frac{1}{2} D_{\para} f^{(o2)} \right] 
     = 0.
\ee
From Eq.~(\ref{eq:oPhi}), it follows that 
\be
\left[ D_{\perp} \Phi_{(o)} \right] = 0. 
\ee 
This is rewritten as 
\bea 
&&{} 
\left[ 
       D_{\perp}\left( \frac{\Phi_{(o)}}{r} \right) 
\right]
   = \frac{1}{2} \kappa \sigma \overline{\left(\frac{\Phi_{(o)}}{r} \right)}. 
\label{eq:ojc1}
\eea 

Second, from Eq.~(\ref{eq:odd1}), we see that
\be
0 = \left[ 
           \epsilon^{ab} D_{b} \left( \frac{{\cal F}_{a}}{r} \right) 
    \right]
  = \frac{1}{r^2} \left[
                        \left( \frac{2}{r} D^{a}r D_{a} - \Box_{2} \right)
       \Phi_{(o)} \right] .
\label{con:o}  
\ee
Using the master equation~(\ref{eq:omaster}), we rewrite Eq.~(\ref{con:o}) as 
\be 
   \left[ \Phi_{(o)} \right] = 0. 
\label{eq:ojc2}
\ee
Eqs.~(\ref{eq:ojc1}) and (\ref{eq:ojc2}) should be imposed on
the solutions of the master equation as the boundary conditions at the bubble
wall. We obtain no new condition from Eq.~(\ref{eq:odd2}). 

Substituting the solutions~(\ref{eq:sol}) into the junction 
condition~(\ref{eq:ojc2}), we get 
\be
\int^{\infty}_{-\infty} dk \psi_{l}(\tau,k)
\left\{
       {\cal A}_{+} - {\cal A}_{-} + {\cal B}_{+} - {\cal B}_{-}
\right\}
        = 0, 
\ee  
where 
\be
{\cal A}_{\pm} 
               := A_{\pm} e^{ ik_{\pm} \zeta_{\pm}}|_{\Sigma},
\quad {\cal B}_{\pm} 
               := B_{\pm} e^{ - ik_{\pm} \zeta_{\pm}}|_{\Sigma}.
\ee 
Here, $e^{ \pm ik_{\pm} \zeta_{\pm}}|_{\Sigma}$ are constant phase factors. 
Multiplying $\int^{\infty}_{-\infty} d\tau \psi^*_{l}(\tau,k')$
and using the orthogonality condition~(\ref{eq:ortho}), we obtain
\be
\left[ {\cal A} \right] + \left[ {\cal B} \right] = 0. 
\label{oCoeff1}
\ee
Similarly, from the junction condition~(\ref{eq:ojc1}), we obtain
\be
\left(
      k - i\frac{\kappa \alpha \sigma}{4} 
\right) 
\overline{{\cal A}}
- \left(
         k + i\frac{\kappa \alpha \sigma}{4} 
\right) 
\overline{ {\cal B}} = 0. 
\label{oCoeff2}
\ee
From the conditions~(\ref{oCoeff1}) and (\ref{oCoeff2}), 
we get consequently the relations of the mode expansion
coefficients, 
\be
{\cal B}_{\pm} 
 = \frac{ - i \mu }{ k + i \mu }
                                {\cal A}_{\pm} 
   + \frac{k}{ k + i \mu } 
                                {\cal A}_{\mp}, 
\label{eq:oddrelations}
\ee
where 
\be  
 \mu := \frac{\kappa \alpha \sigma}{4}. 
\ee 

\subsection{Even modes}
From Eq.~(\ref{eq:delqij}), the functions 
$f_{\para \para}$, $f_{\para}{}^{(e1)}$, $f^{(e0)}$, 
and $f^{(e2)}$ are continuous.
Then 
\be
\left[ X_{\para} \right] = \left[ rf_{\para}{}^{(e1)} - \frac{1}{2}
r^2 D_{\para} f^{(e2)} \right] = 0.
\ee
Similarly, from Eq.~(\ref{eq:Kaa4}), we see that
\be
\left[ X_{\perp} \right] = 0.
\ee
It therefore turns out that the vector $X_{a}$ is continuous quantity.
Then, from Eqs.~(\ref{eq:K}), (\ref{def:alpha}), and 
the definition~(\ref{def:Fab}) of ${\cal F}_{ab}$, we get 
\bea
\left[ 
      {\cal F}_{\para \para }
\right] 
&=& 
\left[ 
      f_{\para \para} - \frac{2}{r}D_{\para}\left(r X_{\para} \right) 
      - 4(K^a_a)X_{\perp} -\frac{1}{2}f^{(e0)} -\frac{1}{2}l(l+1)f^{(e2)} 
\right] 
\non \\
&=& 
     - 4 \left[ K^a_a \right] X_{\perp} 
\non \\  
&=& 
     - 2 \kappa \sigma X_{\perp}, 
\label{condition:FpapaF} 
\\
\left[
      {\cal F}^c_c
\right] 
&=& 
\left[ 
      f^{(e0)} + l(l+1)f^{(e2)} + 4 \frac{D_{\para}r}{r} X_{\para} 
      + 4 (K^a_a) X_{\perp} 
\right] \nonumber \\
&=& 
   4\left[ K^{a}_{a} \right] X_{\perp} 
\non \\ 
\qquad \qquad
&=& 
   2 \kappa \sigma X_{\perp}. 
\label{condition:F}
\eea 
Hence, 
\be 
    \left[{\cal F}_{\para \para}\right] + \left[ {\cal F}^c_c \right] = 0, 
\label{condition:Fpapa} 
\ee 
and equivalently 
\be 
\left[   
      {\cal F}_{\perp \perp}
\right]  
      = 0. 
\label{condition:Fpepe}
\ee 
It follows from Eq.~(\ref{eq:Kaa2}) that
\bea
&&{}
\left[{\cal F}_{\para \perp} \right] = 0. 
\label{condition:Fpape} 
\eea
With the help of Eq.~(\ref{condition:Fpape}) and the relation 
\be
\frac{1}{2}(K^p_p)^2_+ + 2 H^2 = \frac{1}{2}(K^p_p)^2_-, 
\ee
Eq.~(\ref{eq:Kaa3}) yields 
\bea
&&{}
\left[ D_{\perp} {\cal F}^c_c \right] 
  + \left[ 
         (K^p_p) \left( 
                       {\cal F}_{\perp \perp} - \frac{1}{2} {\cal F}^c_c 
                 \right) 
    \right] = 0. 
\label{DFF}
\eea 
By using Eqs.~(\ref{eq:K}), (\ref{def:alpha}), and (\ref{condition:Fpepe}), 
we can rewrite Eq.~(\ref{DFF}) as 
\be
\left[ 
   D_{\perp}{\cal F}^c_c 
\right] 
  = - \frac{H^2}{ \kappa \sigma}
      \left[{\cal F}^c_c \right] 
    - \kappa \sigma 
             \left( 
                  \overline{ {\cal F}_{\perp \perp}} 
                  - \frac{1}{2} \overline{{\cal F}^c_c} 
             \right). 
\label{condition:DF}
\ee
Using Eq.~(\ref{Ein:constraint}), we see that no new constraint is obtained 
from the continuity of $(\del K^{(e)}_{\para})^a_a$. 

Now let us express the junction conditions~(\ref{condition:Fpapa}), 
(\ref{condition:Fpape}), and (\ref{condition:DF}) in terms of 
the master variable $\Phi_{(e)}$.
It is useful to rewrite Eq.~(\ref{exist:pote}) in the form 
\bea
&&{}
{\cal F}_{ab} 
 = r D_{a} D_{b} \left( \frac{\Phi_{(e)}}{r} \right)
 + r \left( \frac{D_{a}r}{r}\right) D_{b} \left( \frac{\Phi_{(e)}}{r} \right)
 + r \left( \frac{D_{b}r}{r}\right) D_{a} \left( \frac{\Phi_{(e)}}{r} \right). 
\label{eq:Fab} 
\eea 

First, by using the even mode master equation~(\ref{eq:emaster}), 
we see that the jump of the trace part of Eq.~(\ref{eq:Fab}) becomes 
\bea
&&{}
\left[{\cal F}^c_c \right] 
 = 2 \cosh \tau 
   \left\{
          \sqrt{1 - H^2 \alpha^2}
    \left( \partial_{\chi}\frac{\Phi_{(e)}}{r} \right)_{+} 
  - \left( \partial_{\chi} \frac{\Phi_{(e)}}{r} \right)_{-}
   \right\} 
           - \frac{1}{\alpha}  
  \left\{ 
         2 \sinh \tau \partial_{\tau} - \frac{l(l+1)}{\cosh \tau} 
  \right\}
\left[\frac{\Phi_{(e)}}{r} \right], 
\label{eq:dF}
\eea
and the jump of ${\cal F}_{\para \para}$, 
\be
\left[ {\cal F}_{\para \para} \right] 
  = 
   \frac{2}{\alpha} 
    \left\{ \cosh \tau \partial^2_{\tau} 
              + \sinh \tau \partial_{\tau} + \frac{l(l+1)}{2 \cosh \tau} 
    \right\}
    \left[ 
           \frac{\Phi_{(e)}}{r}
    \right]. 
\label{eq:dFpp}
\ee
Substituting Eqs.~(\ref{eq:dF}) and (\ref{eq:dFpp}) into 
the condition~(\ref{condition:Fpapa}), we obtain
\bea  
&&{}
\left\{ 
        \partial^2_{\tau} + \frac{l(l + 1)}{ \cosh^2 \tau} 
\right\}
       \left[\frac{\Phi_{(e)}}{r} \right]  
  = 
     - \alpha 
       \left\{ 
              \sqrt{1 - H^2 \alpha^2} 
              \left( \partial_{\chi}\frac{\Phi_{(e)}}{r} \right)_{+}
            - \left(\partial_{\chi} \frac{\Phi_{(e)}}{r} \right)_{-} 
       \right\}.
\label{eq:ejc1}
\eea
              
Next, from Eqs.~(\ref{eq:Fab}) and~(\ref{condition:Fpape}), we obtain 
\be
\partial_{\tau} \left( \cosh \tau \left[ \partial_{\chi}
\frac{\Phi_{(e)}}{r} \right]\right) = 0. 
\label{eq:tauchi0}
\ee
With the help of Eq.~(\ref{eq:emaster}), 
from Eq.~(\ref{eq:Fab}), we get 
\bea
&&{}
\left[ 
  D_{\perp}{\cal F}^c_c  
\right] 
  = \frac{2}{\alpha^2}
    \left\{ \cosh \tau \partial^2_{\tau} + \sinh \tau \partial_{\tau}
    + \frac{l(l + 1)}{2 \cosh\tau}\right\}
    \left\{ \sqrt{1 - H^2 \alpha^2}\left( \frac{\Phi_{(e)}}{r} \right)_{+}
    - \left( \frac{\Phi_{(e)}}{r} \right)_{-}  \right\} 
\non \\
&&{}
\qquad \qquad 
- \frac{2}{\alpha} 
  \left\{ 
     \tanh \tau \partial_{\tau}
         \left( 
             \cosh \tau \left[ \partial_{\chi} \frac{\Phi_{(e)}}{r} \right]
         \right)
     - \frac{(l-1)(l+2)}{2 \cosh\tau}
       \left[ \partial_{\chi} \frac{\Phi_{(e)}}{r} \right] 
  \right\}, 
\label{eq:DperpF}
\\ 
&&{} 
\overline{{\cal F}_{\perp \perp}} 
          - \frac{1}{2} \overline{{\cal F}^c_c} 
  = \frac{1}{\alpha}
 \left\{  \cosh \tau \partial^2_{\tau} + \sinh \tau \partial_{\tau}
+ \frac{l(l + 1)}{2 \cosh\tau} \right\}
\overline{\left(\frac{\Phi_{(e)}}{r} \right)}. 
\label{eq:averageFpp}
\eea
Substituting Eqs.~(\ref{eq:dF}), (\ref{eq:tauchi0}), (\ref{eq:DperpF}), and 
(\ref{eq:averageFpp}) into the condition~(\ref{condition:DF}),
we obtain 
\be
%(l-1)(l + 2)
\left[ \partial_{\chi} \left(\frac{\Phi_{(e)}}{r}\right) \right] = 0.
\label{eq:ejc2}
\ee
Eqs.~(\ref{eq:ejc1}) and (\ref{eq:ejc2}) are the boundary conditions at the 
bubble wall for the even mode master variable $\Phi_{(e)}$.

Now, let us translate the conditions~(\ref{eq:ejc1}) and (\ref{eq:ejc2}) 
into the conditions on the mode expansion coefficients. 
Substituting the solutions~(\ref{eq:sol}) into the conditions~(\ref{eq:ejc1}) 
and (\ref{eq:ejc2}), we obtain, for $k\neq0$,  
\bea  
&&{}
ik \left(\left[{\cal A}\right] + \left[{\cal B}\right] \right)
   = \sqrt{1 - H^2 \alpha^2}
     \left( {\cal A}_{+} - {\cal B}_{+} \right)
     + {\cal A}_{-} - {\cal B}_{-}, \\
&&{}
\overline{{\cal A}} - \overline{{\cal B}} = 0 .
\eea
Then, using the equation  
\be
1 - \sqrt{1 - H^2 \alpha^2} = 2 \mu,
\ee
we finally obtain the relations of the even mode expansion coefficients, 
\be
  {\cal B}_{\pm} 
            = \frac{ i \mu}{ k + i \mu} 
              {\cal A}_{\pm} 
            + \frac{k}{k + i \mu} 
              {\cal A}_{\mp}.
\label{eq:evenrelations}
\ee

Eq.~(\ref{eq:evenrelations}) are the same form as the odd mode case 
Eq.~(\ref{eq:oddrelations}), apart from the difference of the sign of the
first term. They gives the manner of reflection and transmission 
of incident gravitational waves by the bubble wall. 

One expect that the poles of Eqs.~(\ref{eq:oddrelations}) 
and~(\ref{eq:evenrelations}) might correspond to the characteristic modes 
of the gravitational waves emitted by the bubble wall. 
Both the poles, in the present cases, are pure imaginary: $k = - i \mu$. 
The behavior of the pole modes should be examined in the Milne universe 
$t_{-} > r_{-}$ in accordance with the one-bubble open inflation context. 
By the analytic extension of $\Phi$ with respect to the coordinates 
transformation~(\ref{ext:chart}), the metric perturbations ${\cal F}_{ab}$ 
and ${\cal F}_{a}$ in the Milne universe are obtained. 
Then, one can easily observe that ${\cal F}_{ab}$ and ${\cal F}_{a}$ 
corresponding to the pole modes diverge at the center $r_{-} = 0$. 
Hence, the solution $\Phi$ of the pole modes should be excluded 
from the global solutions by the regularity condition at the center. 
For the same reason, $k= 0$ even mode should also be excluded. 
As a result, there is no characteristic mode. 

Since, for odd mode perturbations, there is no variable which describes 
physical deformation of the bubble wall, the result for the odd modes 
is naturally understood. However, for even modes, $X_{\perp}$ expresses 
the physical deformation of the bubble wall~\cite{IsIs}. 
Its behavior is described by the even mode master variables $\Phi_{(e) \pm}$ 
and hence the bubble wall is not a rigid surface. In fact, 
from Eqs.~(\ref{condition:FpapaF}) and (\ref{eq:dFpp}), it is obtained that 
\be 
X_{\perp} 
    = - \frac{1}{ \kappa \alpha \sigma} \left\{ \cosh \tau \partial^2_{\tau}
      + \sinh \tau \partial_{\tau} + \frac{l(l+1)}{2 \cosh \tau} \right\}
        \left[ \frac{\Phi_{(e)}}{r}\right]. 
\label{sol:X}
\ee 
Thus, the  result for the even modes is nontrivial. 
It is concluded that the behavior of the bubble wall is completely 
accompanied with the metric perturbations and the 
bubble wall has no its own dynamical degree of freedom. 

It is straight forward to calculate the case that the interior of the bubble
is also de Sitter spacetime which corresponds to a slow rollover inflation
phase in the one-bubble open inflation scenario.
In this case we can also obtain the same results,
Eqs.~(\ref{eq:oddrelations}) and (\ref{eq:evenrelations}),
only apart from the value of $\alpha$. In this case
\be
 \alpha = \frac{4 \kappa\sigma}{\sqrt{16 (H_{+}^2 - H_{-}^2)^2
        + 8 \kappa^2 \sigma^2 (H_{+}^2 + H_{-}^2) + \kappa^4 \sigma^4}},
\ee
where $\Lambda_{\pm} := 3H^2_{\pm}$ ($H_{+}^2 > H_{-}^2$)
are the cosmological constants of the outside and the inside of the bubble 
(see Appendix C). 

\section{Conclusions and discussion} 
We have analyzed the coupled system of perturbations of gravity and 
the bubble nucleating in de Sitter spacetime. 
We have adopted the thin wall approximation to treat the bubble wall. 
We have solved the perturbed Einstein equations on both sides of 
the bubble and connecting the solutions along the bubble wall 
by using the metric junction formalism. 
Thus we have obtained the global solutions of the perturbed Einstein 
equations in the whole spacetime. 
We concluded that the perturbative motion of the bubble wall is 
completely accompanied with the gravitational perturbations; 
the bubble wall oscillates only while incident gravitational waves 
go across. The result is essentially the same as that in Ref.~\cite{KIF}. 

In the context of the one-bubble open inflation scenario, bubble wall 
perturbations are investigated by several authors~\cite{Qstate}. 
Recently, Tanaka and Sasaki have shown that, 
once the gravitational perturbations are taken into account, 
the bubble wall fluctuation modes disappear~\cite{tama}. 
Our result is consistent with their work. 

The behavior of the wall clarified in the present work is different from
the intuitive expectation such that deformed bubble wall oscillates 
by its own dynamics and the oscillation damps gradually by emission of 
gravitational waves. 
As shown in Ref.~\cite{GVG}, the perturbations on the wall can be described 
by a single scalar field $\phi$ on the hypersurface which represents 
the background motion of the wall: three-dimensional de Sitter spacetime 
in the present case. 
When we ignore the gravitational back reaction on the motion of 
the bubble wall, we obtain the perturbed equation of motion for 
the bubble wall as the Klein-Gordon wave equation on the background 
hypersurface, 
\be 
\left( 
      \Box_3- m^2 
\right) \phi = 0, 
\label{eq:KG}
\ee 
where $\Box_3$ is the d'Alembertian on the three-dimensional 
background hypersurface 
and the mass term $m^2$ is expressed by $H^2$ and $\sigma$. 
This equation has oscillatory solutions and the non-spherical oscillations 
of the wall become a source of gravitational waves. 
Thus, if the influence of the gravity might be negligible, the solution of 
Eq.~(\ref{eq:KG}) have described the behavior of the wall well. 
In the present case, however, we see the qualitatively different behavior of 
the wall motion, taking the back reaction of the gravitational perturbations 
into account. Our results suggest that the back reaction 
of the gravitational perturbations on the motion of a domain wall 
should not be ignored. 

The difference of our results and usual analysis
may appear in the equation of motion for the bubble wall. 
By using the junction condition, we can easily obtain the equation
of perturbative motion of the bubble wall coupled with the gravitational 
perturbations as the equation for $X_{\perp}$ on the orbit space.
From the junction condition~(\ref{eq:JCK}), we get 
\be
\del \overline{K} = 0.
\ee
Then substituting Eqs.~(\ref{eq:Kaa1}), (\ref{eq:Kaa3}), and (\ref{eq:Kaa4}) 
into this condition, we obtain the perturbed equation of motion for 
the wall as
\bea
&&{}
\left\{
       - D^2_{\para} - 2 \left( \frac{D_{\para}r}{r}\right) D_{\para}
       - \frac{l(l+1)}{r^2} + \frac{3}{2}H^2
       + 3 \left( \frac{H^2}{\kappa \sigma} \right)^2
       +\frac{3\kappa^2 \sigma^2}{16}
\right\} X_{\perp} 
\non \\
&&{}
 + \frac{3 H^2}{2 \kappa \sigma}{} 
 \left(
       \overline{{\cal F}_{\perp \perp}} - \frac{1}{2}\overline{{\cal F}^c_c}
 \right) 
- \frac{1}{2r^4}
  {} \overline{ 
                D_{\para}\left( 
                               r^4 {\cal F}_{\para \perp}
                         \right) 
               }
- \frac{1}{4} \overline{D_{\perp}{\cal F}^c_c} = 0. 
\label{eq:eomX}  
\eea  
Comparing this equation with Eq.~(\ref{eq:KG}), we notice that
there are additional terms described by the metric perturbations. 
The task which we should do is solving the coupled system of 
the perturbed Einstein equations and the equation of motion for 
the bubble wall~(\ref{eq:eomX}) simultaneously. 
This work has been accomplished by using the metric junction formalism 
in the present paper. Consequently, the quantities of the second line 
in Eq.~(\ref{eq:eomX}) are comparable to the quantities of the first line, 
i.e., the back reaction of the gravitational waves on the bubble wall motion 
is not negligible. 

The equation of motion~(\ref{eq:eomX}) is obtained from the junction condition 
which is derived from the Einstein equations. 
It is natural because the Einstein equations contain the equation of motion 
for matters. 
We can also derive the perturbed equation of motion~(\ref{eq:eomX}) 
directly from the Nambu-Goto action for the domain wall. 
It will appear in Ref.~\cite{IsIs}.

It is interesting to study the quantum fluctuations of the bubble wall 
in the early universe. 
Garriga and Vilenkin investigated the quantization of the scalar field 
satisfying Eq.~(\ref{eq:KG})~\cite{GVQuantum}. 
However, if we consider a domain wall coupled with gravitational waves, 
the dynamical freedom of the wall is lost. Thus, another quantization scheme 
for the coupled system may be applied.

One of the reasons for the result obtained in the present paper may be 
the fact that the stress-energy tensor is proportional to the intrinsic 
metric of the wall hypersurface (see Eq.~(\ref{eq:stress})). 
More general cases are under investigation. 

\section*{Acknowledgments}
We are grateful to Dr. T.~Tanaka for informative comments and 
fruitful discussion. We would like to thank Dr. K.~Nakamura 
for helpful and useful discussion. 
We would also like to thank Professor A. Hosoya for his continuous 
encouragement. 

%%%%%%%%%%%%%%%%%%%%%%%%%%%%%%%%%%%%%%%%%%%%%%%%%%%%%%%%%%%%%%%%%%%%%%%%
\appendix
%%%%%%%%%%%%%%%%%%%%%%%%%%
\section{Tensor harmonics}
Here we give the definitions and the basic properties of 
the tensor harmonics. We denote the metric of the unit two-sphere by 
\be 
   \Omega_{pq} dz^p dz^q = d\theta^2 + \sin^2 \theta d\phi^2, 
\ee 
and the Laplace-Beltrami operator by 
$\hat{\triangle}_{2} := \Omega^{pq}\hat{D}_{p}\hat{D}_{q}$. 

The bases of the tensor harmonics are chosen so that each has a definite 
parity with respect to the transformation 
$ 
(\theta, \phi) \rightarrow (\pi - \theta, \phi + \pi) 
$ 
which maps each point on the unit sphere to its antipodal point. 
Under this choice the harmonics $T$ which transform as 
\be 
T \longrightarrow (-1)^l T
\ee 
are called even, and those which transform as 
\be 
T \longrightarrow (-1)^{l+1} T 
\ee 
are called odd. 

\subsection{Scalar harmonics}
The scalar harmonics are the spherical harmonic functions $Y_{lm}(z^p)$
which satisfy the equations
\bea
&&{}
\left\{\hat{\triangle}_{2} + l(l+1) \right\} Y_{lm}= 0, 
\\
&&{}
\partial_{\phi} Y_{lm} = i m Y_{lm}. 
\eea 
They are expressed in terms of the Legendre functions as 
\be 
Y_{lm}(\theta,\phi) = \sqrt{\frac{(2l + 1)(l-m)!}{4\pi (l+m)!}}
P^m_l (\cos \theta)e^{im\phi}. 
\ee 
These are all even. 

\subsection{Vector harmonics}
The vector harmonics on the unit sphere are defined by 
\bea
&&{} (V^{(e)}{}_{lm})_p := \hat{D}_p Y_{lm}
\quad \mbox{for even modes},
\\
&&{} (V^{(o)}{}_{lm})_p := \epsilon_{pq}\hat{D}^q Y_{lm}
\quad \mbox{for odd modes}, 
\eea
where $\epsilon_{pq}$ is the two-dimensional Levi-Civit\'a antisymmetric
tensor. 
Hereafter, the subscripts $(e)$ and $(o)$ express the even 
and the odd parity, respectively. 
Any vector on the unit sphere can be expanded by these vector harmonics.
By definition, the vector harmonics satisfy 
\bea
&&{}
\hat{D}_p (V^{(e)}{}_{lm})^{p} = -l(l+1)Y_{lm}, 
\\
&&{}
\hat{D}_p (V^{(o)}{}_{lm})^{p} = 0, 
\\
&&{}
\hat{D}_{[p}(V^{(e)}{}_{lm})_{q]} = 0, 
\\
&&{}
\hat{D}_{[p}(V^{(o)}{}_{lm})_{q]} = \frac{1}{2} l(l+1)\epsilon_{pq}Y_{lm},
\\
&&{}
\left( \hat{\triangle}_{2} + l^2 + l - 1 \right) (V_{lm})_{p} = 0 
\quad
\mbox{for both the even and the odd modes}, 
\eea

The vector harmonics vanish for $l=0$.

\subsection{Tensor harmonics}
Any second lank tensor field on the unit sphere can be expanded by
the tensor harmonics which are defined by 
\bea
&&{} 
(T^{(e0)}{}_{lm})_{pq} := \frac{1}{2} \Omega_{pq} Y_{lm} , 
\\
&&{} (T^{(e2)}{}_{lm})_{pq} 
     := \left\{
               \hat{D}_p \hat{D}_q + \frac{1}{2}l(l+1) \Omega_{pq} 
        \right\}Y_{lm} , 
\\
&&{} (T^{(o2)}{}_{lm})_{pq} 
     := \frac{1}{2}(\epsilon_{qr}\hat{D}_p + \epsilon_{pr}\hat{D}_q ) 
        \hat{D}^{r}Y_{lm}.  
\eea  
The harmonic tensor $(T^{(e0)}{}_{lm})_{pq}$ is 
essentially of scalar type and satisfies 
\bea
&&{}
\left\{ 
       \hat{\triangle}_{2} + l(l+1) 
\right\} (T^{(e0)}{}_{lm})_{pq} = 0, 
\\
&&{}
(T^{(e0)}{}_{lm})_{p}{}^p = Y_{lm}, 
\\
&&{}
\hat{D}_q (T^{(e0)}{}_{lm})^{qp} 
              = \frac{1}{2}(V^{(e)}{}_{lm})^p. 
\\
\eea
The harmonics $(T^{(e2)}{}_{lm}){}_{pq}$ and $(T^{(o2)}{}_{lm}){}_{pq}$ are
purely tensorial and satisfy 
\bea
&&{}
\left\{ 
       \hat{\triangle}_{2} + l^2 + l - 4 
\right\} (T^{(e2,o2)}{}_{lm}) {}_{pq} = 0, 
\\
&&{}
(T^{(e2,o2)}{}_{lm}){}_p{}^p = 0, 
\\
&&{}
\hat{D}_q (T^{(e2,o 2)}{}_{lm}){}^{qp} 
              = - \frac{1}{2} (l - 1)(l + 2) (V^{(e,o)}{}_{lm}){}^p.
\eea

\section{Derivation of the even mode master equation} 
We show that the even mode perturbed Einstein equations~(\ref{Ein:constraint}) 
and (\ref{Ein:evolve}) reduce to the master equation~(\ref{eq:emaster}). 

Substituting Eq.~(\ref{exist:pote}) into Eq.~(\ref{Ein:evolve}), 
we get 
\be 
    \left( 
           D_a D_b + H^2 \gamma_{ab} 
    \right) L = 0, 
\label{eq:L} 
\ee 
where 
\be 
    L := \left\{ 
                 - r^2 \Box_{2} + 2 r \left(D^{c} r \right) D_{c} + (l-1)(l+2) 
         \right\} \Phi_{(e)}. 
\ee 
Regarding Eq.~(\ref{eq:L}) as a second order differential equation 
for $L$, we obtain the general solution of Eq.~(\ref{eq:L}), 
denoted by $L_{\rm s}$, as 
\be 
   L_{\rm s}
      = C_1 r + \sqrt{1 - H^2 r^2} 
                        \left( C_2 \sinh Ht +  C_3 \cosh Ht \right), 
\ee   
where $C_1$, $C_2$, and $C_3$ are arbitrary constants. 
Thus, $\Phi_{(e)}$ should satisfy 
\be 
\left\{ 
       - r^2 \Box_{2} + 2 r \left(D^{c} r \right) D_{c} + (l-1)(l+2) 
\right\} \Phi_{(e)}
 = L_{\rm s}. 
\ee  
By the transformation of $\Phi_{(e)}$ such as 
\be 
    \Phi_{(e)} \longrightarrow 
    \Phi'_{(e)} := \Phi_{(e)} - \frac{L_{\rm s}}{(l - 1)(l + 2)}  
               + \frac{2 C_{1}r}{(l-1)l(l+1)(l+2)} ,  
\label{transf:phi}
\ee  
we obtain 
\be 
   L' := 
        \left\{ 
               - r^2 \Box_{2} + 2 r \left(D^{c} r \right) D_{c} + (l-1)(l+2) 
        \right\} \Phi'_{(e)}
      = 0 .   
\label{eq:L'}
\ee 
The transformation~(\ref{transf:phi}) does not change ${\cal F}_{ab}$. 
Thus, rewriting $\Phi'_{(e)}$ by $\Phi_{(e)}$, we can reduce 
Eq.~(\ref{eq:L'}) to the equation for $\Phi_{(e)}$ of the form 
\bea 
     \left( 
           - \Box_{2} + \frac{l(l+1)}{r^2} 
     \right)
            \left(\frac{\Phi_{(e)}}{r}\right) = 0. 
\eea 

\section{de Sitter - de Sitter case}
Let us consider the case that the inside of the bubble is also 
de Sitter spacetime, where the cosmological constants of the outside~(inside) 
are $\Lambda_{+(-)} \equiv 3H_{+(-)}^{2}$ ($H_{-}^{2} < H_{+}^{2}$). 
From Eq.~(\ref{Kpq}), the $(p,q)$ components of the extrinsic curvatures are 
\be 
(K^{q}_{p})_{\pm} 
        = - \frac{\sqrt{ 1 - H^2_{\pm} r_{\pm}^2 
          + \dot{r}_{\pm}^2}}{r_{\pm}} \del^q_p. 
\label{dede:Kpq}
\ee
Substituting Eq.~(\ref{dede:Kpq}) into the junction 
condition~(\ref{eq:JCKij}), 
we obtain the equation of the bubble wall motion 
\be
1 + \dot{r}^2 = \left( \frac{r}{\alpha} \right)^2, 
\ee  
where 
\be
\alpha = \frac{4 \kappa \sigma}{\sqrt{ 16 (H^2_{+} - H^2_{-})^2
+ 8 \kappa^2 \sigma^2 (H^2_{+} + H^2_{-}) + \kappa^4 \sigma^4}}.
\ee  
The solution is
\be
r = \alpha \cosh \tau,
\ee 
where $\tau$ is normalized by $\alpha$. 
This conforms to the ${\rm O(3,1)}$ symmetry of the bubble wall. 

As in the Minkowski--de Sitter case, we can consider the metric perturbations 
of the geometry in terms of the gauge-invariant variables introduced in 
the section~3. 
In the hyperbolic charts defined by Eq.~(\ref{Charts:Hyp}) with 
\bea 
&&{} 
a(\chi_{\pm}) := \frac{1}{H_{\pm}}\sin H_{\pm}\!{\chi_{\pm}}, 
\quad 
a(\chi_{\pm})|_{\Sigma} = \alpha, 
\eea 
we obtain the master equations~(\ref{eq:omaster}) and (\ref{eq:emaster}) 
for the master scalar $\Phi$ and solutions 
\be  
\left(\frac{\Phi}{r} \right)_{\pm}
    = \int_{-\infty}^{\infty} dk \psi_l(\tau,k)
      \left\{ 
             A_{\pm}(k) e^{i k \zeta_{\pm}} + B_{\pm}(k) e^{- i k \zeta_{\pm}}
      \right\},
\ee
where
\be
\zeta(\chi_{\pm}) := \mp \ln\left\{ \tan\left( \frac{H_{\pm}\chi_{\pm}}{2}
\right) \right\}. 
\ee  

The odd mode junction conditions for $\Phi_{(o)}$ are described as 
\bea
&&{}
\left[ D_{\perp}\left( \frac{\Phi_{(o)}}{r} \right)\right]
     = \frac{1}{2}\kappa \sigma \overline{\left(\frac{\Phi_{(o)}}{r}\right)}, 
\\
&&{} 
\left[ \frac{\Phi_{(o)}}{r} \right] = 0. 
\eea
These are the same forms as the Minkowski--de Sitter case.
In terms of the expansion coefficients $A_{\pm}$ and $B_{\pm}$
of the solution, these conditions reduce to 
\be
{\cal B}_{\pm} 
=
 - \frac{ i \mu}{ k + i \mu} 
           {\cal A}_{\pm} 
  + \frac{k}{k + i \mu} 
           {\cal A}_{\mp}.
\label{dede:oddrelation}
\ee

For even modes, the junction conditions of the metric perturbation expansion 
coefficients are 
\bea
&&{}
      \left[
            {\cal F}^c_c 
      \right] = 2 \kappa \sigma X_{\perp}, 
\label{a} 
\\
&&{}
      \left[ 
            {\cal F}_{\para \para} 
      \right] = - 2 \kappa \sigma X_{\perp}, 
\label{b} 
\\
&&{}
      \left[
           {\cal F}_{\perp \perp}  
      \right] 
= 0, 
\label{c} 
\\
&&{}
      \left[
            {\cal F}_{\para \perp} 
      \right] = 0, 
\label{d} 
\\
&&{}
\left[ 
    D_{\perp}{\cal F}^c_c 
\right] 
= 
- \frac{H^2_{+} - H^2_{-}}{\kappa \sigma} 
    \left[ 
         {\cal F}^c_c 
    \right] 
- \kappa \sigma 
  \left(
       \overline{{\cal F}_{\perp \perp}} - \frac{1}{2} \overline{{\cal F}^c_c}
  \right). 
\label{e}
\eea 
Rewriting these conditions in terms of the even mode master 
variable~$\Phi_{(e)}$ and combining them, we obtain the following constraints 
for $\Phi_{(e)}$, 
\bea 
&&{} 
\left\{ 
       \partial^2_{\tau} + \frac{l(l+1)}{\cosh^2 \tau} 
\right\}
        \left[\frac{\Phi_{(e)}}{r} \right]
= 
 - \alpha \left\{ 
                  \sqrt{1 - H^2_{+}\alpha^2} 
                         \left( 
                               \partial_{\chi}\frac{\Phi_{(e)}}{r}
                         \right)_{+}
                - \sqrt{1 - H^2_{-}\alpha^2} 
                         \left( 
                               \partial_{\chi} \frac{\Phi_{(e)}}{r} 
                         \right)_{-} 
          \right\}= 0, 
\label{ejc1}
\\ 
&&{} 
(l - 1)(l + 2) 
      \left[ 
            \partial_{\chi} \frac{\Phi_{(e)}}{r} 
      \right] = 0. 
\label{ejc2}
\eea 
These constraints yield the following relations of the mode expansion 
coefficients 
\be
{\cal B}_{\pm} 
       = \frac{ i \mu}{ k + i \mu} 
               {\cal A}_{\pm} 
       + \frac{k}{k + i \mu} 
               {\cal A}_{\mp}.
\label{dede:evenrelation}
\ee
Eqs.~(\ref{dede:oddrelation}) and (\ref{dede:evenrelation}) are 
the same forms as Eqs.~(\ref{eq:oddrelations}) and (\ref{eq:evenrelations}), 
respectively. 

\newpage
%

%%%%%%%%%%%%%%%%%%%%%%%%%%%%%%%%%%%%%%%%%%%%%%%%%%%%%%%%%%%%%%%%%%%%%%%%
\newpage
\begin{figure}
 \centerline{\epsfxsize= 10.0cm \epsfbox{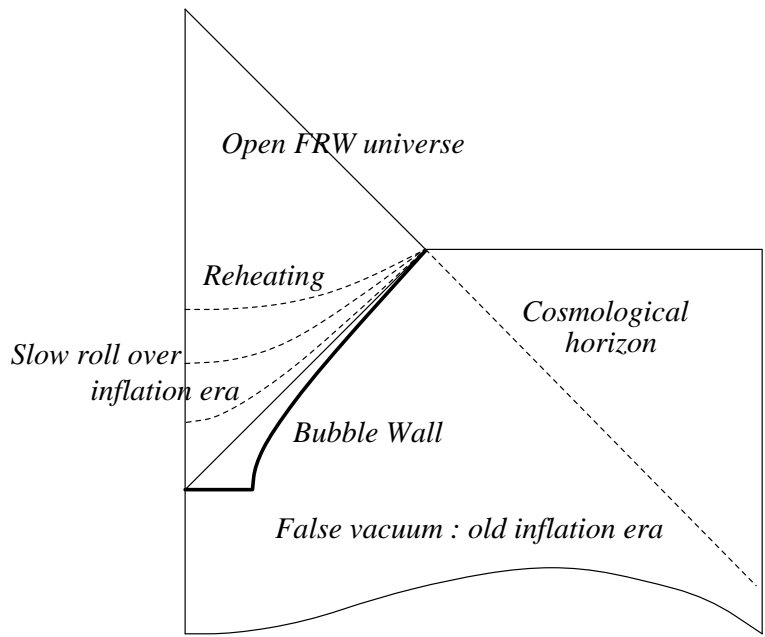}}
        \caption{The Penrose diagram of the one-bubble open inflation
                 universe. }
        \protect \label{fig:Oboi}
\end{figure}
%%%%%%%%%%%%%%%%%%%%%%%%%%
\begin{figure}
 \centerline{\epsfxsize= 9.0cm \epsfbox{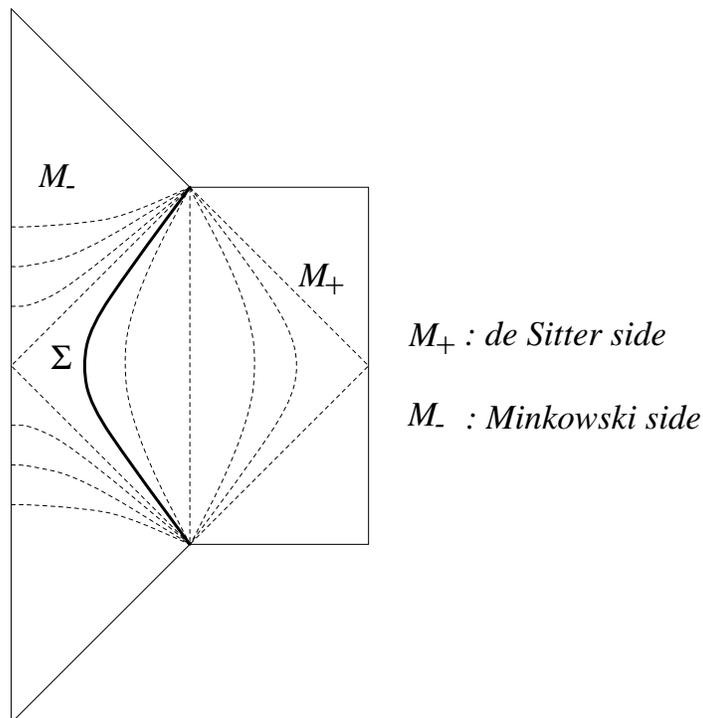}}
        \caption{The Penrose diagram of the background geometry.
                 Minkowski spacetime $\lq\lq M_{-}"$ and de Sitter 
                 spacetime $\lq\lq M_{+}"$ are connected at the bubble wall 
                 $\Sigma$ (thick line). The timelike dashed lines denote 
                 $\chi = \mbox{const.}$ hypersurfaces and the spacelike 
                 dashed lines in $M_{-}$, $\eta = \mbox{const.}$ 
                 hypersurfaces, which correspond to open FRW time-slices. }
        \protect \label{fig:MdS}
\end{figure}
%%%%%%%%%%%%%%%%%%%%%%%%%%
\begin{figure}
 \centerline{\epsfxsize= 4.5cm \epsfbox{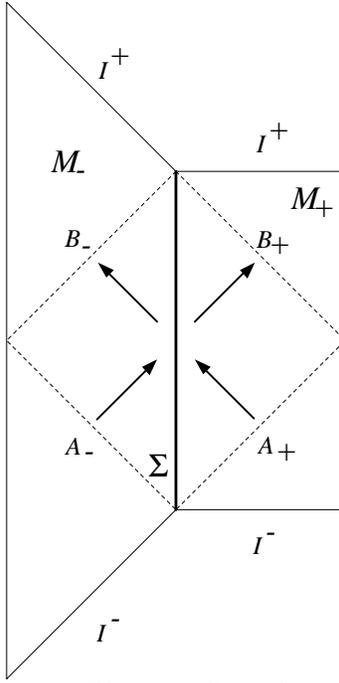}}
  \caption{ The block enclosed by dashed lines, 
            the Rindler horizons, is covered by the hyperbolic  
            (the spherical Rindler) charts. 
            The thick line represents the bubble wall $\Sigma$.
            The mode expansion coefficient $A_{+}(A_{-})$ represents 
            the wave propagating from the past Rindler horizon 
            in $M_{+}(M_{-})$ toward the bubble wall $\Sigma$. 
            Similarly, $B_{+}(B_{-})$ represents the wave 
            from $\Sigma$ toward the future 
            Rindler horizon in $M_{+}(M_{-})$. }
   \protect \label{fig:Waves}
\end{figure}
%%%%%%%%%%%%%%%%%%%%%%%%%%

\end{document}